\documentclass[a4paper,11pt]{article}
\usepackage{jheppub} % for details on the use of the package, please see the JINST-author-manual
\usepackage{lineno}
\usepackage{mathrsfs,amsmath, amscd, amsthm,amssymb, amsfonts, verbatim,subfigure, enumerate,slashed}
\usepackage{tikz,tikz-cd}

\def\U{\ensuremath{\mathrm{U}}}
\def\SU{\ensuremath{\mathrm{SU}}}

\def\cM{\mathcal M}\def\cN{\mathcal N}\def\cO{\mathcal O}

\def\CC{\mathbb C}

\def\RR{\mathbb R}

  \def\fp{\mathfrak p}

\def\R{\mathbb{R}} 
\def\C{\mathbb{C}}

\DeclareMathOperator{\Spec}{Spec}

\DeclareMathOperator{\Spin}{Spin}

\DeclareMathOperator{\Sym}{Sym}

\def\bu{\bullet}

\newtheorem{thm}{Theorem}[section]

\newtheorem*{thm*}{Theorem}

\theoremstyle{definition}

\newtheorem{dfn/lem}[thm]{Definition/Lemma}

\theoremstyle{remark}
\newtheorem{rmk}[thm]{Remark}

\def\d{{\rm d}}

\arxivnumber{2602.22318} % if you have one

\title{\boldmath Twisting BFSS \& IKKT}

\author{Fabian Hahner, Natalie M. Paquette}
\affiliation{University of Washington,  Department of Physics \\ 3910 15th Ave NE, Seattle, WA 98195,  U.S.A}

\emailAdd{fhahner@uw.edu}
\emailAdd{npaquett@uw.edu}

\abstract{In this note we initiate the study of ``twisted holography'' for the dualities involving the BFSS matrix quantum mechanics and the IKKT matrix model in their $N \rightarrow \infty$ limits. We identify the admissible twists of each model, compute their cohomology in the BV-BRST formalism, and identify them---in the planar limit and in perturbation theory around the trivial background---with corresponding twists of IIA and IIB string theories, respectively. The twisted gravitational duals make manifest certain infinite dimensional symmetry algebras. In the BFSS example, the dual IIA supergravity twists are also obtained as certain zero mode truncations of the minimal (1/16-BPS) and maximal (1/4-BPS) twists of eleven-dimensional supergravity.  }

\begin{document}

\setcounter{tocdepth}{1}

\maketitle
\flushbottom

\setlength{\parskip}{7pt}

\section{Introduction}
Twisted holography, as introduced by \cite{CostelloLi, Costello:2018zrm, Costello:2020jbh}, is the study of supersymmetry protected subsectors of holographic dualities. It recasts such observables as mathematical objects, and exploits isomorphisms among these structures. Homological and homotopical algebra have played a particularly important role, arising from their centrality in 1.) the BV-BRST formalism of (string) field theories and 2.) open-closed duality of topological string theories. Moreover, recent progress in understanding twisted supergravity \cite{CostelloLiTypeI, CostelloMtheory2, spinortwist, RSW, MaxTwist, CY2, sconf, HPR} has shown that these theories are best expressed as holomorphic-topological field theories, with action functionals based on geometric objects. 

There are two equivalence classes of twists of eleven-dimensional supergravity which are possible in a flat background \cite{RSW, MaxTwist}; they are called the minimal and maximal twists. This opens the door for explorations in a version of twisted holography for the duality of Banks-Fischler-Shenker-Susskind (BFSS) \cite{BFSS}, which states that scattering in M-theory in eleven-dimensional asymptotically flat spacetime can be computed using $U(N)$ D0 matrix quantum mechanics (MQM) in the $N \rightarrow \infty$ limit. 

As a first step towards understanding a twisted version of this proposed duality, we identify two (equivalence classes of) square-zero supercharges in the MQM and find that they correspond to twists of IIA supergravity that are subsectors of the maximal and minimal twists of eleven-dimensional supergravity. We similarly classify twists of the Ishibashi-Kawai-Kitazawa-Tsuchiya (IKKT) matrix model \cite{IKKT} and compare them with twists of IIB supergravity. 

In more detail, we find that the gauge-invariant observables of the minimal twist of the IKKT model in the $N \rightarrow \infty$ limit can be matched with minimally twisted type IIB supergravity \cite{CostelloLi} via standard techniques of twisted holography. The minimal twist of IIB supergravity is given by a version of Bershadsky-Cecotti-Ooguri-Vafa (BCOV) \cite{BCOV} theory on $\mathbb{C}^5$, which is the closed string field theory of the B-model topological string and perturbatively governs complex structure deformations of the target space. Similarly, we find a correspondence between the minimally twisted BFSS MQM and the $\SU(4)$-twist of type IIA supergravity, which can be understood as the closed string field theory of a mixed A/B-model topological string. 

In addition, we find that non-minimal twisting supercharges in both the BFSS and IKKT model lead to twisted theories with acyclic (i.e. perturbatively trivial) BV complexes, signaling that all interesting structure for these twists is encoded in the global structure of field space\footnote{This phenomenom is roughly similar to what happens in the Donaldson--Witten twist of the four-dimensional $\cN=2$ gauge multiplet.}. In the case of the BFSS model, non-trivial observables can still be constructed via topological descent. We match these twists with twists of type IIB and type IIA supergravity that have not been explored much in the literature so far. They can be understood from the spacetime point of view as deformations of BCOV theory, these by linear superpotentials, which indeed render the theory trivial in perturbation theory.

The BV-BRST formalism, roughly, equips the space of fields (including all ghosts, antifields, etc.) with the structure of an $L_{\infty}$ algebra. Since the physical degrees of freedom are encoded in the cohomology with respect to the BV differential, it is often useful to pass to \textit{quasi-isomorphic} models of the same classical field theory; a quasi-isomorphic model is one which has the same underlying cohomology as the original model, and hence the same physical observables, but may inherit nontrivial $L_{\infty}$ brackets. In particular, we make frequent use of quasi-isomorphic versions of the twisted supergravity theories called their ``minimal models'', which are quasi-isomorphic $L_{\infty}$ algebras that have vanishing differentials, and are often convenient to work with. As a vector space, the minimal model is simply the cohomology of the original theory. In particular, the minimal models are isomorphic to the infinite-dimensional symmetry algebras (possibly up to a central extension) of the twisted theories. By constructing isomorphisms between twisted IIA/IIB supergravity and, respectively, twisted BFSS/IKKT, we can explicate the infinite dimensional symmetry enhancements of the 1/4 and 1/16-BPS subsectors of the physical theories in the planar limit. By analogy to more familiar examples of twisted holography \cite{Costello:2018zrm, Costello:2020jbh, FPW}, we expect these algebras are large enough to fix the large-$N$ BPS 3-point functions.

\subsection*{Acknowledgments}
We thank K. Costello, S. Raghavendran, I. Saberi, and B. Williams for enjoyable discussions on this and other closely related topics over the years. Special thanks are due to S. Raghavendran for helpful comments on the draft. NP and FH are both supported by the DOE Early Career Research Program under award DE-SC0022924. NP is also supported by funds from the Department of Physics and the College of Arts \& Sciences at the University of Washington, and the Simons Foundation as part of the Simons Collaboration on Celestial Holography.

\section{Twisted holography for the IKKT matrix model}

\subsection{Review of the IKKT matrix model}

The IKKT matrix model~\cite{IKKT} can be obtained by dimensionally reducing ten-dimensional super Yang-Mills theory to a point. Accordingly, its physical degrees of freedom are ten adjoint-valued matrices, $X \in V \otimes \mathfrak{u}(N)$, that transform in the vector representation $V$ of $\Spin(10)$ (now viewed as R-symmetry) together with sixteen fermions $\psi^\alpha \in S_+ \otimes \mathfrak{u}(N)$ that live in the positive chirality spin representation. 

The action functional takes the form
\begin{equation} \label{eq: action-ikkt}
    S = -\frac{1}{g} \mathrm{tr} \left(  \frac{1}{4} [X^I , X^J] [X_I, X_J] - \frac{1}{2} \psi \Gamma^I [X_I, \psi] \right).
\end{equation}
The theory is acted upon by the super Lie algebra $\fp_{IKKT} = \mathfrak{so}(10) \ltimes \Pi S_+$, where $\Pi S_+$ denotes the odd abelian subalgebra generated by the supercharges. Concretely, the supersymmetry transformations of the fields take the form
\begin{equation} \label{eq: susy0-ikkt}
    \delta X^I = \varepsilon \Gamma^I \psi \qquad \text{and} \qquad \delta \psi = \varepsilon \Gamma_{IJ} [X^I,X^J].
\end{equation}
Working in the BV formalism, we extend the space of fields by a ghost for gauge transformations $c \in \mathfrak{u}(N)$, as well as antifields $X^*_I$, $\psi^*$ and $c^*$, which will implement the equations of motion cohomologically. We can encode both gauge and supersymmetry transformations in the BV action using standard methods (see e.g. \cite{JurcoReview} for a modern review),
\begin{equation}
    S_{BV} = S_0 + S_{\mathrm{gauge}} + S_{\mathrm{susy}},
\end{equation}
which can also be obtained from the BV action of ten-dimensional super Yang--Mills theory via dimensional reduction. Let us briefly describe the different terms that appear. First, all fields simply transform in the adjoint representation of the gauge group, so that we have
\begin{equation}
    S_{\mathrm{gauge}} = \mathrm{tr}(c^* [c,c] + X^{*I} [c, X_I] + \psi^* [c, \psi] ).
\end{equation} The variations of the antifields implement the gauge transformations of their respective partner fields. 
The full structure of the supersymmetry transformations must include their closure terms up to gauge transformations and the equations of motion\footnote{In mathematical terms, these closure relations signal that the complex of BV fields does not form a Lie module for the supersymmetry algebra, but only an $L_\infty$ module.} and is encoded by
\begin{equation}\label{eq: ikkt-susy}
    S_{\mathrm{susy}} = \mathrm{tr}( X^{*}_I \varepsilon \Gamma^I \psi + \frac{1}{2} \psi^* \varepsilon \Gamma_{IJ} [X^I, X^J] + c^* (\varepsilon \Gamma_I \varepsilon)X^I + (\varepsilon \Gamma^I \varepsilon) (\psi^* \Gamma^I \psi^*) - 2(\varepsilon \psi^*)^2 ).
\end{equation}
Recall that the supersymmetry transformations can generally be understood by coupling to a supergravity background and then restricting the bosonic ghosts for the supersymmetry to constant values. In the above equation, we identify $\varepsilon$ as this bosonic ghost. This perspective is particularly useful in relation to twisted supergravity~\cite{CostelloLi} since the twisting procedure amounts to working in a background where $\varepsilon$ has a non-zero expectation value. We note that the terms linear in $\varepsilon$ of~\eqref{eq: ikkt-susy} correspond to the ordinary supersymmetry transformations~\eqref{eq: susy0-ikkt}, while the quadratic terms represent higher order terms in the $L_\infty$ module structure. These terms have the following origins. First, the supersymmetry transformations only close up to a gauge transformation with the field dependent parameter $\varepsilon \Gamma^I \varepsilon X_I$. This is signaled by the term proportional to the antifield of the ghost. Second, the supersymmetry equations only close up to the equation of motion of the fermion, which is encoded in the final terms that are quadratic in $\psi^*$.

As usual, the BV differential is induced from the BV action via the antibracket $Q_{BV} = \{S_{BV} , - \}_{BV}$, where $\{F, G \}_{BV} := \int d^dx\,
\left(
\frac{\overrightarrow{\delta} F}{\delta \Phi^i(x)}
\frac{\overleftarrow{\delta} G}{\delta \Phi_i^*(x)}
-
\frac{\overrightarrow{\delta} F}{\delta \Phi_i^*(x)}
\frac{\overleftarrow{\delta} G}{\delta \Phi^i(x)}
\right)$ in terms of left and right derivatives with respect to the fields and antifields. We note that it decomposes into three different pieces
\begin{equation}
    Q_{BV} =  \delta_{\mathrm{e.o.m}} + \delta_{\mathrm{gauge}} + \delta_{\varepsilon}
\end{equation}
corresponding to the three different terms in the BV action. These implement the equations of motion, gauge invariance and supersymmetry transformations, respectively.

Let us explicitly state the full supersymmetry transformations that arise from the BV action.
\begin{equation} \label{eq: susy trafos}
    \begin{array}{cc}
        \delta c = \varepsilon \Gamma^I \varepsilon X_I & \delta X^*_I = \varepsilon \Gamma_{IJ} [\psi^* , X^J] + \varepsilon \Gamma_I \varepsilon c^* \\[6pt]
        \delta X^I = \varepsilon \Gamma^I \psi & \delta \psi^* = \varepsilon \Gamma^I X^*_I \\[6pt]
        \delta \psi = \frac{1}{2} \varepsilon \Gamma_{IJ} [X^I,X^J] + (\varepsilon \Gamma^I \varepsilon) \Gamma_I \psi^* - 2\varepsilon (\varepsilon \psi^*) &   \delta c^* = 0
    \end{array}
\end{equation}

\begin{rmk}
    We remark that the transformations in~\eqref{eq: susy trafos} have an algebro-geometric interpretation in the following way. The cone of pure spinors in $S_+$ forms an affine variety cut out by ten quadratic equations. Resolving the ring of functions of this variety in free $\cO(S_+)$-modules gives a cochain complex whose resolution differential precisely takes the form of the above supersymmetry transformations. This can be seen by means of the pure spinor superfield formalism~\cite{perspectives}.
\end{rmk}

\subsubsection{Relation to IIB string theory}
The IKKT matrix model was conjectured to give rise to a non-perturbative description of type IIB string theory~\cite{IKKT} (see also \cite{Ciceri:2025maa}). Various analytical and numerical explorations of this proposal have been performed \cite{Kimura:2000ur, Kitazawa:2002vh, Aoki:1998bq, Nishimura:2020blu, Anagnostopoulos:2022dak}. We remark that there are still many questions regarding this duality\footnote{There are also recent works that explore the duality for the polarized IKKT theory, which is better behaved, see e.g. \cite{Hartnoll:2024csr, Komatsu:2024ydh, Komatsu:2024bop}.}, particularly in Lorentzian signature. Since twisting captures supersymmetry protected quantities in Euclidean signature, we explore only a small well-defined piece of the original proposal. Moreover, note that twisting as it is formalized in this note is adapted to work in perturbation theory around a chosen solution to the BV field equations. In other words, the BV complexes we write and act upon with twisting supercharges are local models of a formal neighborhood in field space around the chosen background solution. As we will see below in more detail, these models are not immediately sensitive to the global structures of field space that include non-perturbative effects like the existence of other non-trivial saddle points of the action functional (e.g. instantons). For a more global approach to the BV formalism, see for example~\cite{Elliott:2015rja, Alfonsi:2023qpv}.

In terms of type IIB string theory, the IKKT model arises as the worldvolume theory describing the dynamics of $N$ interacting D(-1)-branes. Type IIB string theory, as well as its low energy approximation type IIB supergravity, are theories with an action of the ten-dimensional $\cN=(2,0)$ supersymmetry algebra that features 32 supercharges. Our above description of the IKKT model as the dimensional reduction of ten-dimensional super Yang--Mills theory makes only 16 of these apparent. These 16 supercharges are called the \emph{dynamical} supercharges. The remaining 16 supercharges, called \emph{kinematical} supercharges, arise as symmetries of the center of mass degrees of freedom of the D(-1)-branes. As such, they only act via shifts on the trace parts of the fermion field $\psi$. In the following, we will choose twisting supercharges that always act as dynamical supersymmetries; in work to appear~\cite{HP2} we will explicate twists that include the kinematical SUSYs, and provide a 1-to-1 match with the twists of IIB and IIA supergravities recently classified in \cite{iib-nv, iia-nv}.

\subsection{Twisting supercharges}
The moduli space of twists\footnote{In the context of twisting, we work with complexified (super) Lie algebras. In the following, whenever we discuss twists of the IKKT and later the BFSS model, we work with both complexified versions of the supersymmetry algebra and the gauge algebras. Once symmetry groups act on spacetime (e.g. as subgroups of the Lorentz group), one typically chooses the real structure corresponding to Euclidean signature.} for a supersymmetric field theory with supersymmetry algebra $\fp = \fp_+ \oplus \Pi \fp_-$ (decomposed here into its even and odd generators) is the variety of odd, square-zero elements~\cite{NV,ElliottSafronov},
\begin{equation}
    Y = \{ Q \in \fp_- \: | \: [Q,Q]=0 \}.
\end{equation}
Here and throughout we use the convention of graded Lie brackets, i.e. $[x,y]
=-(-1)^{|x||y|}[y,x]$, with $|x|, |y| \in \mathbb{Z}/2\mathbb{Z}$ denoting fermion parity.
If $\fp$ is a super Poincar\'e algebra, this space decomposes into orbits under the Lorentz group and R-symmetry. In $\fp_{IKKT}$ all odd elements are square zero, so that the nilpotence variety is all of the affine space $S_+$. It decomposes into two orbits~\cite{Igusa}:
\begin{itemize}
    \item[---] A minimal orbit of dimension 11. In this orbit, the stabilizer is semidirect product of $SU(5)$ with the additive group of antisymmetric rank-two tensors on $\mathbb{C}^5$: $\mathrm{SU(5) \ltimes \wedge^2 \CC^5}$.
    \item[---] The open orbit of dimension 16. In this orbit, the stabilizer is $\Spin(7) \ltimes \CC^8$.
\end{itemize}
We note that the twisting supercharges can be distiguished by their vector invariants. The minimal twisting supercharge is a pure spinor inside $S_+$; in other words, it has a vanishing vector invariant
\begin{equation}
    \varepsilon^{(0)} \Gamma^I \varepsilon^{(0)} = 0.
\end{equation}
On the other hand, the maximal twisting supercharge has non-vanishing vector invariant.

The variety of square-zero elements in the type IIB supersymmetry algebra and its orbit classification was described in~\cite{iib-nv}. Based on these results, it is natural to expect that the rank one twists of type IIB supergravity can be matched by these two twists of the IKKT matrix model. In order to match the rank two twists, a non-trivial interplay between the dynamical supersymmetries and the kinematical supersymmetries is required \cite{HP2}. 

\subsection{The minimal twist of the IKKT matrix model}
Let us start by investigating the minimal twist. We will obtain the twisted theory in two different ways. First, we describe it as a dimensional reduction of holomorphically twisted ten-dimensional super Yang--Mills theory. Then, we compute the twist directly from the BV complex of IKKT.

\subsubsection{The minimal twist as a dimensional reduction} \label{sec: min-tw-ikkt}
The holomorphic twist of ten-dimensional super Yang--Mills theory is holomorphic Chern--Simons theory on $\CC^5$. The BV complex of fields organizes into Dolbeault forms on $\mathbb{C}^5$ valued in the gauge group
\begin{equation}
    (\Omega^{0,\bullet}(\CC^5) , \bar{\partial}) \otimes \mathfrak{gl}_N
\end{equation}
so that for $\alpha \in \Omega^{0,\bullet}(\CC^5) \otimes \mathfrak{gl}_N$ the BV action is
\begin{equation}
    S_{BV}[\alpha] = \int \Omega \wedge \mathrm{tr} \left( \alpha \wedge ( \bar{\partial} \alpha + \frac{2}{3}[\alpha, \alpha]) \right) 
\end{equation}
where $\Omega = \d z_1 \wedge \dots \wedge \d z_5$ is the holomorphic volume form. This twist was first considered by Baulieu~\cite{BaulieuCS}; see also~\cite{CY4} for an extensive discussion.

Dimensionally reducing this theory to a point leaves us with a collection of five odd variables $\theta^i$ transforming in the fundamental representation of $\mathrm{SU}(5)$. Thus, the BV fields are simply,
\begin{equation}
   A_N = \CC[\theta^1, \dots, \theta^5] \otimes \mathfrak{gl}_N.
\end{equation}
This result is a special case of Lemma 7.1.1 in~\cite{CostelloLi}. For an element $\alpha \in A_N$, the BV action is just $\int \d^5 \theta \mathrm{tr}(\alpha [\alpha, \alpha])$.
\subsubsection{Direct computation of the minimal twist}
To compute the minimal twist directly from the BV fields of the IKKT model, we decompose them under the $\SU(5)$ stabilizer of the twisting supercharge.
\begin{equation}
    X^I \mapsto (Z^i , \bar{Z}_i) \qquad \psi \mapsto (\psi^{(0)}, \psi^{(2)}, \psi^{(4)}).
\end{equation}
Here, $Z^i$ with $i=1 \dots 5$ transforms in the fundamental representation of $\SU(5)$, while $\bar{Z}_i$ is in the antifundamental. The fermion decomposes according to the identification
\begin{equation} \label{eq: spinor-decomp}
    S_+ \cong \wedge^0 \CC^5 \oplus \wedge^2 \CC^5 \oplus \wedge^4 \CC^5
\end{equation}
and the upper index denotes form degree. We have a similar decomposition for the antifields, which live in the corresponding dual representations; in particular we take the convention such that $\psi^*$ decomposes into $(\psi^{*(1)} , \psi^{*(3)}, \psi^{*(5)})$.

We now fix a minimal twisting supercharge $Q = \varepsilon^\alpha Q_\alpha$. This can be achieved by setting the zero-form component $\varepsilon^{(0)} = 1$ and all other components to zero. In order to twist, we assign this value to the parameters $\varepsilon$ in the BV action, i.e. we specialize to
\begin{equation}
    S_{BV}[\varepsilon^{(0)} = 1, \varepsilon^{(2)} = \varepsilon^{(4)} = 0]    
\end{equation}
Note that, compared to the background with $\varepsilon=0$, this precisely deforms the BV differential by the action of the twisting supercharge $Q$ to give
\begin{equation} \label{eq: tw-diff}
    \delta_{\mathrm{e.o.m.}} + \delta_{\mathrm{gauge}} + \delta_Q. 
\end{equation}
We now investigate the BV theory described by this deformed BV differential.

The differential~\eqref{eq: tw-diff} features two different types of terms. Terms linear in the BV fields originate from $\delta_Q$ and lead to the formation of acyclic pairs of fields, which cancel once we take cohomology. This is natural, since a theory capturing only BPS physics of the original theory should  have fewer degrees of freedom than its parent. 

Further, there are terms of higher order in the BV fields coming from all three pieces of the differential; these give rise to the interactions of the twisted theory. Accordingly, we proceed in a two-step procedure: first, we identify all acyclic pairs from the linear transformations and then move on to identify what terms the non-linear transformations induce.

Let us denote the linear terms of the supersymmetry transformations as $\delta_Q^{(1)}$. They decompose as follows.
\begin{equation}
\begin{array}{ll}
    % Left Column & Right Column
    \delta_Q^{(1)} c = 0 & \delta_Q^{(1)} Z^*_i = 0 \\[6pt]
    \delta_Q^{(1)} Z^i = 0 & \delta_Q^{(1)} \bar{Z}^{*i} = 0 \\[6pt]
    \delta_Q^{(1)} \bar{Z}_i = \varepsilon_{ijklm} \psi^{(4)jklm} & \delta_Q^{(1)} \psi^{*(1)i} = \bar{Z}^{*i} \\[6pt]
    \delta_Q^{(1)} \psi^{(0)} = \varepsilon_{ijklm} \psi^{*(5)ijklm} \quad & \delta_Q^{(1)} \psi^{*(3)} = 0 \\[6pt]
    \delta_Q^{(1)} \psi^{(2)} = 0 & \delta_Q^{(1)} \psi^{*(5)} = 0 \\[6pt]
    \delta_Q^{(1)} \psi^{(4)} = 0 & \delta_Q^{(1)} c^* = 0
\end{array}
\end{equation}
From here, we can immediately read off the cohomology. It is spanned by $(c, Z^i, \psi^{(2)}, \psi^{*(3)} , Z^*_i , c^*)$. We can compare to the description we obtained by dimensional reduction. Explicitly, we identify fields with polynomials in $\theta$ as recorded in the following table. Note that the parity obtained from $\theta$-degree in this table is related to the physical parity by a shift\footnote{$\theta$-degree naturally corresponds to the degree of the underlying dg Lie algebra, which is related to the ghost degree in the BV formalism by a shift of 1. This is a generalization of the more familiar fact that ordinary gauge transformations are bosonic, whereas the ghosts representing them in the BRST formalism are fermionic.}; in terms of physical parity, $c$ is fermionic, $Z^i$ is bosonic, and so on.
\begin{table}[h]
\centering
\renewcommand{\arraystretch}{1.5} % Increases row height to make the math formulas easier to read
\begin{tabular}{|c|c|c|c|c|c|}
\hline
$(\theta^i)^0$ & $(\theta^i)^1$ & $(\theta^i)^2$ & $(\theta^i)^3$ & $(\theta^i)^4$ & $(\theta^i)^5$ \\
\hline
$c$ & $Z^i$ & $\psi^{(2)}$ & $\psi^{*(3)}$ & $Z^*_i$ & $c^*$ \\
\hline
\end{tabular}
\caption{Fields of minimally twisted IKKT.} \label{t: ikkt}
\end{table}
We now include the non-linear terms from the differential. Together, these give rise to the bracket on $\CC[\theta^1, \dots, \theta^5] \otimes \mathfrak{gl}_N$. For example, the quadratic supersymmetry transformation on the fermion $\psi^{(2)}$,
\begin{equation}
    \delta_Q (\psi^{(2)})^{ij} \sim [Z^i , Z^j],
\end{equation}
corresponds to the binary bracket formed by multiplying terms linear in $\theta$ and applying the Lie bracket in $\mathfrak{gl}_N$. Terms from gauge transformations,
\begin{equation}
    \delta_{\mathrm{gauge}} X = [c,X] \quad, \quad \delta_{\mathrm{gauge}} \psi = [c,\psi] \quad \text{etc.}
\end{equation}
correspond to multiplication by elements in theta-degree zero. Finally, the equations of motion give rise to
\begin{equation}
    \delta_{\mathrm{e.o.m}} Z^*_i = \varepsilon_{ijklm} [\psi^{(2) jk} , \psi^{(2) lm}]  \qquad \delta_{\mathrm{e.o.m.}} \psi^{*(3) ijk} = [Z^i, \psi^{(2) jk}] ,
\end{equation}
that correspond to multiplications of elements in theta-degree two as well as one and two respectively.

Further, it is now easy to see how the BV action of the IKKT model restricts to the cubic action obtained by dimensionally reducing holomorphic Chern--Simons theory above. Explicitly, we have
\begin{equation}
    S_{BV} = \varepsilon_{ijklm}\mathrm{tr}( [Z^i , \psi^{(2)jk}] \psi^{(2)lm} + [Z^i , Z^j] \psi^{* (3)klm}) + \mathrm{tr}([c,c] c^* + [c,Z^i] Z^*_i + [c,\psi^{(2)}] \psi^{*{3}})
\end{equation}
We note that the first two terms originate from $S_0$ and $S_{\mathrm{susy}}$ respectively, while the remaining terms come from $S_{\mathrm{gauge}}$.

\subsection{Twisted holography for minimally twisted IKKT} \label{sec: hol-ikkt}
In the spirit of the relation between the IKKT model and type IIB string theory, one expects a similar relation between minimally twisted IKKT and a twisted version of type IIB. To make this precise, we may apply theorems from homological algebra which formalize open-closed string duality. 

For any commutative differential graded algebra $A$, the Loday-Quillen-Tsygan (LQT) theorem establishes a quasi-isomorphism\footnote{A quasi-isomorphism is the relevant notion of equivalence when studying BV-BRST theories. They are morphisms of chain complexes which induce isomorphisms in cohomology. Therefore, it will often be enlightening to pass to quasi-isomorphic models, as when computing free or projective resolutions of quadratic algebras.}
\begin{equation}
    \Sym^\bullet(CC^\bullet(A)[-1]) \longrightarrow C^\bullet(A \otimes \mathfrak{gl}_\infty).
\end{equation}
On the right hand side, we have the Chevalley-Eilenberg cochains of a Lie algebra $A \otimes \mathfrak{gl}_{\infty}$, where $\mathfrak{gl}_{\infty}$ is the $N \rightarrow \infty$ limit of the gauge algebra\footnote{At finite $N$, the LQT map fails to be a quasi-isomorphism due to the familiar trace relations.}. This is nothing but the BRST complex for gauge invariants, treated mathematically as a cochain complex equipped with the usual BRST differential. 

Restricting the right side to single trace operators, one obtains a quasi-isomorphism to cyclic cochains of $A$, $CC^\bullet(A)[-1]$, which we will define in Remark 2.2 below. The left hand side is a free commutative algebra, the symmetric algebra, on cyclic cochains; the latter should be understood as constructing traces of algebra words around closed loops, respecting cyclic ordering. In the spirit of open-closed string duality, the cyclic cohomology is identified with a (twisted) closed string sector whose low energy description is given by (twisted) supergravity (see for example~\cite{Costello:2018zrm,GGHZ} for discussions). The $[-1]$ denotes a shift of elements upwards by one cohomological degree; this ensures that ghost number zero operators in gauge theory map to higher ghost number operators in gravity, consistent with the nonexistence of local gauge-invariant operators. 

\begin{rmk}
    Let us briefly recall the definition of cyclic cohomology and its relation to string theory. For original references on cyclic cohomology, see~\cite{Connes, Quillen}; for the relation to physics, e.g. \cite{Kapustin:2004df, Aspinwall:2009isa} and in particular string field theory \cite{Moeller:2010mh}. Let $A$ be a differential graded algebra with internal differential $d$. A cyclic $n$-cochain on $A$ is a graded linear map $\phi: A^{\otimes (n+1)} \longrightarrow \mathbb{C}$ satisfying the cyclic symmetry $\phi(a_0, \dots, a_n) = (-1)^n \epsilon \phi(a_n, a_0, \dots, a_{n-1})$, where $\epsilon$ denotes the Koszul sign. The complex of cyclic cochains is equipped with the differential
    \begin{equation}
        b\phi(a_0, \dots, a_{n+1}) = \sum_{i=0}^n (-1)^i \phi(a_0, \dots , a_i a_{i+1}, \dots , a_{n+1}) + (-1)^{n+1}\phi(a_{n+1}a_0, a_1, \dots , a_n),
    \end{equation}
    where Koszul signs have been suppressed for brevity. Because $A$ is a dg-algebra, the cyclic complex also carries a differential $d_*$ induced by the action of $d$ on the arguments. The cyclic cohomology is defined as the cohomology of the total differential $D = b + d_*$. 

    Crucially, at degree $n=0$, a cyclic cochain is simply a linear functional $\tau: A \longrightarrow \mathbb{C}$, and the closure condition $b\tau = 0$ reduces to the trace property $\tau(a_0 a_1) = \tau(a_1 a_0)$. Cyclic cohomology thus can be viewed as the homological extension of the space of traces on $A$.

    In string theory, open string states and their interactions are governed by a differential graded algebra~\cite{Witten:1985cc}, where the internal differential $d$ corresponds to the BRST operator. Geometrically, evaluating a trace corresponds to identifying the string endpoints, which produces a closed string. When computing the coupling of a bulk closed string state to boundary open string insertions on a disk, the $S^1$-diffeomorphism symmetry of the disk boundary enforces cyclical symmetry among the insertions. Combining the cyclic differential with the BRST operator one finds that the spectrum of physical closed string states is described by the cyclic cohomology of the open string dg-algebra \footnote{At the worldsheet level, one first computes Hochschild cochains, which are disc one-point-functions with one closed string vertex operator insertion and an arbitrary number of open string operators on the disc boundary. To obtain on-shell closed string states, one then computes semirelative cohomology, such that the $b_0^{-}$-ghost annihilates physical states, which enforces the residual cyclic gauge symmetry on the disc.}. As such cyclic cohomology describes the closed string sector as consistent deformations of the open string field theory.
\end{rmk}

Let us now apply this framework to the twisted IKKT matrix model. Here, we have $A = \CC[\theta^1, \dots, \theta^5]$ and the cyclic cohomology of $A$ can be evaluated using the following standard procedure (see for example~\cite{Costello:2018zrm}). 
First, we compute the Hochshild cohomology of $A$ via the Hochschild--Kostant--Rosenberg theorem. This theorem identifies the Hochschild cohomology of $A$ with polyvector fields\footnote{Recall that the space of polyvector fields on a complex manifold $X$ is defined as $\mathrm{PV}^{i,j}(X) = \Omega^{0,i}(X, \wedge^i T^{(1,0)} X )$, which are Dolbeault forms valued in powers of the holomorphic tangent bundle.} on $\Spec(A)$, the geometric space whose functions are given by $A$. In our case, we have
\begin{equation}
    HH^\bullet(A) \cong \mathrm{PV} (\CC^{0|5}) = \CC[\theta^i, \partial_{\theta^i}].
\end{equation}
Here, the differentials $\partial_{\theta^i}$ are now even (i.e. commuting variables). Upon completion, we can identify the Hochschild homology with holomorphic polyvector fields on $\CC^5$
\begin{equation}
    \mathrm{PV}_{hol}^\bullet(\CC^5) = \cO(\CC^5) [\partial_{z_1} , \dots \partial_{z_5}],
\end{equation}
where the $\partial_{z_i}$ are now odd variables. 
This is a special case of the more general statement that polyvector fields on Koszul dual algebras are quasi-isomorphic. However, here this can be directly seen by mapping
\begin{equation} \label{eq: Koszul}
    \theta^i \mapsto \partial_{z_i} \quad \text{and} \quad \partial_{\theta^i} \mapsto z_i .
\end{equation} This Koszul duality mapping, whereby fermionic generators are swapped with spacetime derivatives, is the mathematical mechanism for how a point brane ``knows'' about its ambient geometry, and can also be understood through the existence of universal, gauge-invariant couplings between operators on the brane worldvolume and closed string fields (see \cite{PW} for a review\footnote{More concretely, a bulk-brane coupling is associated to a Maurer-Cartan element (via the descent procedure), which produces an $L_{\infty}$ morphism from the bulk algebra to the cyclic cochains governing endomorphisms or deformations of the brane theory.}). 

We can now apply a Dolbeault resolution to holomorphic polyvector fields to find the quasi-isomorphic complex
\begin{equation}
    (\mathrm{PV}^{\bullet, \bullet}(\CC^5) , \bar{\partial}) = (\Omega^{0,\bullet}(\CC^5 , \wedge^\bullet T_{\CC^5}^{(1,0)}) ,\bar{\partial}).
\end{equation}
As a final step, we relate Hochschild to cyclic cohomology by the Connes spectral sequence. The complex on the first page of this spectral sequence is $(HH^\bullet(A)[\![t]\!] , tB)$, where $t$ is a variable in homological degree two and $B$ is the Connes operator. In our case $B$ is the divergence operator $\partial_{\Omega}$ on polyvector fields. Further, the spectral sequence abuts at the first page so that we can identify
\begin{equation}\label{eq: fullBCOV}
    CC^\bullet(A) \cong (\mathrm{PV}^{\bu, \bu}(\CC^5)[\![t]\!] \: , \: \bar{\partial} + t \partial_\Omega).
\end{equation}
This is the BV complex of fields of BCOV theory on $\CC^5$ in the formulation of Costello and Li~\cite{Costello:2012cy, CLBCOV1}. In this description of BCOV theory, the nonlocality in the original formulation~\cite{BCOV} has been resolved by introducing the parameter $t$ and comes at the price of the BV pairing being degenerate.\footnote{Recall that fields in the original formulation of BCOV theory \cite{BCOV} are restricted to the kernel of $\partial$ in polyvector fields and the action functional has a kinetic term of the form $\int \alpha \bar{\partial} \partial^{-1} \alpha$.}

Interactions can be included in the correspondence by equipping both sides with suitable brackets. The fields of BCOV theory are naturally equipped with the Schouten bracket, making it into a dg Lie algebra. This is mirrored on the gauge theory side, where the BV antibracket acts on gauge invariant observables. Polyvector fields with the Schouten bracket describes BCOV theory as a degenerate BV theory; in this formulation the existence of an action functional is prevented by the non-degenerate nature of the pairing. 

Alternatively, one can apply a non-polynomial field redefinition to express the interactions in a form that is perhaps more natural from the perspective of string field theory. After this field redefinition, the interaction term for $\alpha = \sum_k t^k \alpha_k$ reads
\begin{equation}
    I[\alpha] = \sum_{n \geq 3} \frac{1}{n!} I_n[\alpha] \qquad I_n[\alpha]= \sum_{k_1 + \dots + k_n = n-3} C_{k_1 \dots k_n} \int \Omega \wedge (\Omega \vee \alpha_1 \wedge \dots \wedge \alpha_{k_n}).
\end{equation}
Here, $\Omega = \d z_1 \wedge \ldots \wedge \d z_5$ is the holomorphic volume form. The coefficients $C_{k_1 \dots k_n}$ are fixed as genus zero Gromov--Witten invariants 
\begin{equation}
    C_{k_1 \dots k_n} = \int_{\bar{\cM}_{0,n}} \psi^{k_1} \dots \psi^{k_n} = \frac{(n-3)!}{k_1! \dots k_n!}.
\end{equation}
Finally, we can also simply undo the resolution of the constraint $\ker(\partial)$ by taking cohomology. This recovers the original non-local action functional of BCOV theory.

Since the identification between the twisted gauge theory and gravity side involves several steps, let us summarize the chain of relations in the following diagram.
\begin{equation} \label{eq: tw-hol}
    \begin{tikzcd}[row sep=large, column sep=large]
    (HH^\bullet(A)[\![t]\!], tB) \arrow[r, "\text{Connes}"] & 
    CC^\bullet(A)[-1] \arrow[r, "\text{LQT}"] & 
    C^\bullet_{\substack{\text{single} \\ \text{trace}}}(A \otimes \mathfrak{gl}_\infty) \\
    (\mathrm{PV}(\mathbb{C}^{0|5})[\![t]\!], t\partial) \arrow[u, "\text{HKR}" right] & 
    (\mathrm{PV}_{hol}(\mathbb{C}^5), t\partial_\Omega) \arrow[l, "\substack{\text{Koszul} \\ \text{Duality}}" below] & 
    (\mathrm{PV}^{\bullet,\bullet}(\mathbb{C}^5)[\![t]\!], \bar{\partial} + t\partial_\Omega) \arrow[l, "\bar{\partial}\text{-cohom}" below]
\end{tikzcd}
\end{equation}
The bottom right diagram corresponds to the BV complex of BCOV theory on $\CC^5$. It is related to the cyclic cochains of $A$ by first taking cohomology and thereby restricting to holomorphic polyvector fields, then applying Koszul duality (explicitly given by~\eqref{eq: Koszul}). The HKR theorem is then used to convert a polyvector field into a Hochschild cochain of $A$. Here, a map is given by
\begin{equation}\label{eq:HKR}
    f(\theta) \partial_{\theta_{i_1}} \dots \partial_{\theta_{i_k}} \mapsto \left( (a_1, \dots a_k) \mapsto \frac{1}{k!} \sum_\sigma f(\theta) \partial_{\theta_{i_1}} (a_{\sigma_1}) \dots \partial_{\theta_{i_k}}(a_{\sigma_k}) \right)
\end{equation}
where the sum is over all permutations and the divergence operator on polyvector fields maps to the Connes B-operator. The next step is applying the Connes spectral sequence; for details we refer to~\cite{Loday-cyclic}. Finally, the LQT map constructs a single trace operator out of a cyclic cochain $\phi: A^{\otimes (n+1)} \longrightarrow \CC$ by assigning
\begin{equation}
    \cO_\phi(a_0, \dots, a_n ) = \mathrm{tr}(\phi(a_0, \dots, a_n)) \qquad a_i \in A.
\end{equation}
To provide a concrete illustration, consider a constant bivector field $\Pi = \partial_{z_i} \wedge \partial_{z_j} \in \mathrm{PV}^{2,0}(\CC^5)$. This is the twisted remnant of the B-field on the closed string side \cite{SW, Jurco:2013upa}. Under Koszul duality, it maps to vector field $\theta^i \theta^j \in \mathrm{PV}^0(\CC^{0|5})$. The HKR isomorphism \eqref{eq:HKR} identifies this directly with the Hochschild 0-cocycle $\theta^i \theta^j \in A$. Converting this to a cyclic 0-cochain via the canonical volume form $\int d^5\theta$ gives the map $a \mapsto \int d^5\theta \, a \, \theta^i \theta^j$. For a non-zero integral, $a$ must be the complementary degree-3 element $\epsilon_{i j k l m}\theta^k \theta^l \theta^m$, which corresponds to the fermion antifield $\psi^{*(3)}_{klm}$ in Table~\ref{t: ikkt}. Thus, via the LQT map, this bulk bivector is dual to the single-trace operator $\mathrm{tr}(\psi^{*(3)}_{klm})$. This is natural, since the minimal BV action includes the term $\epsilon_{ijklm}Tr([Z_i, Z_j]\psi^{*(3)}_{klm})$, and hence a constant bivector deformation induces a constant noncommutativity parameter in the $i, j$ directions: $[Z_i, Z_j] = 1$, as expected of a B-field.

BCOV theory enjoys an infinite dimensional Lie algebra of gauge symmetries that preserve the trivial, vacuum solution to the equations of motion around flat space. It is simply given by the divergence-free holomorphic polyvectors (i.e. holomorphic volume form-preserving polyvectors), with Lie bracket the Schouten bracket. Ordinary divergence-free vector fields form a subalgebra denoted by $\mathrm{Vect}_0(\mathbb{C}^5)$, and the bracket reduces to the usual Lie bracket on vector fields.

BCOV theory on $\CC^5$ is the (conjectural) holomorphic twist of type IIB string theory in a flat background~\cite{CostelloLi}. Work addressing this conjecture has been carried out in the supergravity approximation. The holomorphic twist of type IIB supergravity was computed in the free limit in~\cite{spinortwist} using pure spinor superfield methods. It was shown to reproduce a version of \emph{minimal BCOV theory}.
Most straightforwardly, minimal BCOV theory is described by a subcomplex of~\eqref{eq: fullBCOV}. This subcomplex is given by:
\begin{equation} \label{eq: minimal BCOV}
    \left( \bigoplus_{i+j \leq 4} t^i \mathrm{PV}^{j,\bullet}(\CC^5) \: , \: \bar{\partial} + t \partial_\Omega \right).
\end{equation}
The fields in this description of BCOV theory are best viewed as field strengths. Indeed, the physical origin of many of these fields are precisely the field strengths of higher-form gauge fields in supergravity. Relating~\eqref{eq: minimal BCOV} to the twist of IIB supergravity obtained in~\cite{spinortwist} involves choosing potentials for two of the fields. 

It is expected that the difference between full and minimal BCOV theory corresponds to the twist of the infinite tower of massive higher string modes that are absent in the supergravity approximation. After the holomorphic twist, these higher modes no longer propagate. Indeed, minimal BCOV theory is the smallest subcomplex of full BCOV theory that contains all propagating fields (see also the discussion in~\cite[\S5]{RSW}).

One rather direct way to see the infinite dimensional symmetry algebras of twists comes from taking the cohomology of the cochain complexes of BV fields~\eqref{eq: fullBCOV} or~\eqref{eq: minimal BCOV} and performing homotopy transfer. This process gives an $L_\infty$ algebra known as the minimal model of the theory. The infinite-dimensional gauge symmetries then sit as a subalgebra in the minimal model. In~\cite{SuryaYoominimal}, the minimal model for minimal BCOV theory on $\CC^5$ was identified as a one dimensional central extension of $\mathrm{SHO}(\CC^{5|5})$, which is the derived subalgebra of super-divergence free vector fields on $\C^{5|5}$ that at the same time preserve the standard odd symplectic form. In particular, we note that the infinite-dimensional Lie algebra of divergence-free vector fields on $\CC^5$ sits inside these as a subalgebra.

Finally, we remark that from the perspective of topological string theory, the situation for the IKKT model corresponds to the pure B-model. Here, the category of branes is given by coherent sheaves on $\CC^5$ and we recover the minimal twist of the IKKT model as the self-Ext of functions on a point in $\CC^5$.

\subsection{Non-minimally twisted IKKT} \label{sec: spin7-ikkt}
Let us now turn to the second choice of twisting supercharge. This twist is $\Spin(7)$-equivariant and does not originate as a dimensional reduction of twisted ten-dimensional super Yang-Mills theory since the twisting supercharge is not square-zero in the ten-dimensional $\cN=1$ super Poincar\'e algebra. We will find that the cohomology with respect to the twisting supercharge is acyclic; in other words this twist is completely trivial when computed in the formal neighborhood of a point on field space (in the same way as the Donaldson--Witten twist of the four-dimensional $\cN = 2$ vector multiplet is locally trivial in field space\footnote{By contrast, \cite{Cushing:2023rha} construct a version of the twisted supergravity which can couple to the DW twist of SYM and which captures the global moduli stack, rather than just the neighborhood of a single point, and recovers the Cartan model of Diff$^+$ equivariant cohomology.}).
\subsubsection{Direct computation of the non-minimal twist}
We start by decomposing the fields under $\Spin(7)$. Let us first remark on the relevant embedding of $\Spin(7)$ into $\Spin(10)$ that stabilizes our twisting supercharge. For this purpose, let us first restrict the symmetry
\begin{equation}
    \Spin(10) \longrightarrow \Spin(8) \times \Spin(2).
\end{equation}
Under this restriction, the 16 component spinor representation $S_+$ of $\Spin(10)$ decomposes into a direct sum of a positive and negative chirality spin representation
\begin{equation}
    S_+ \longrightarrow \textbf{8}_s \oplus \textbf{8}_c.
\end{equation}
By triality, there are now three different embeddings of $\Spin(7)$ into $\Spin(8)$ that either stabilize a vector in $\textbf{8}_v$, a positive chirality spinor in $\textbf{8}_s$ or a negative chirality spinor in $\textbf{8}_c$. We are using the embedding stabilizing a positive chirality spinor; this gives the following chain of subgroups:
\begin{equation}
    \Spin(7) \times \U(1) \subset \Spin(8) \times \Spin(2) \subset \Spin(10).
\end{equation}
Accordingly, the spinor and vector representations of $\Spin(10)$ decompose as
\begin{equation}
    \begin{split}
        V &\longrightarrow \textbf{8}^0 \oplus \textbf{1}^{+2} \oplus \textbf{1}^{-2} \\
        S_+ &\longrightarrow \textbf{1}^{+1} \oplus \textbf{7}^{+1} \oplus \textbf{8}^{-1}.
    \end{split}
\end{equation}
Here, the superscript denotes the charge under $\Spin(2) \cong \mathrm{U}(1)$. The twisting supercharge lives in $\textbf{1}^{+1}$.
We decompose the field contents accordingly,
\begin{equation}
        X^I \mapsto (X^a, Z_{-2}, \bar{Z}_{+2}) \qquad \psi_\alpha \mapsto (\psi_+, \psi^i_+ , \psi^a_-),
\end{equation}
where $i=1 \dots 7$ is an index for the vector and $a=1 \dots 8$ for the spinor representation. Note that the antfields carry opposite $\U(1)$-charges compared to the fields, so e.g. $X^*_I$ decomposes to $(X^*_a, Z^*_{+2} , \bar{Z}^*_{-2})$.

We restrict to the twisting supercharge to lie in the trivial representation $\textbf{1}^{+1}$ inside $S_+$. Crucially, we note that this time the vector invariant of the twisting supercharge is non-vanishing,
\begin{equation}
    Q_{\text{max}} \Gamma^I Q_{\text{max}} \neq 0.
\end{equation} Restricting the supersymmetry transformations to the twisting supercharge and only writing linear terms, we find the following decomposition.
\begin{equation} \label{eq: delta-spin7}
\begin{array}{ll}
    % Left Column & Right Column
    \delta_Q^{(1)} c = Z_{-2} & \delta_Q^{(1)} Z^*_{+2} = c^*  \\[6pt]
    \delta_Q^{(1)} Z_{-2} = 0 & \delta_Q^{(1)} \bar{Z}^*_{-2} = 0 \\[6pt]
    \delta_Q^{(1)} \bar{Z}_{+2} = \psi_+ & \delta_Q^{(1)} X^*_a = 0 \\[6pt]
    \delta_Q^{(1)} X^a = \psi^a_- \qquad \qquad& \delta_Q^{(1)} \psi^*_- = \bar{Z}^*_{-2} \\[6pt]
    \delta_Q^{(1)} \psi_+ = 0 & \delta_Q^{(1)} \psi^{*}_{a+} = X^*_a \\[6pt]
    \delta_Q^{(1)} \psi^i_+ = \psi^{*i}_- & \delta_Q^{(1)} \psi^*_{i-} = 0 \\[6pt]
    \delta_Q^{(1)} \psi^a_- = 0 & \delta_Q^{(1)} c^* = 0
\end{array}
\end{equation}
Hence, the deformed complex of BV fields is completely acyclic and the twist is locally trivial in field space. In particular, this means that the cohomology of local operators in non-minimally twisted IKKT is empty and that the BV action $S_{BV}$ is exact under the non-minimal twisting supercharge so that it can be written as $S_{BV} = Q_{max} \Psi$. While this establishes the existence of such a $\Psi$, its explicit form can be derived in a systematic way order by order in the following way. First, let $h^{(1)}$ be a contracting homotopy for the linearized differential $\delta_Q^{(1)}$, in other words an operator that undoes the linearized $Q$-variation \eqref{eq: delta-spin7}. Note that this is achieved by simply inverting all non-zero terms in the decomposition~\eqref{eq: delta-spin7}:
if the linearized
twisted differential pairs two fields as
\begin{equation}
    \delta_Q^{(1)} u = v, \qquad \delta_Q^{(1)} v = 0,
\end{equation}
then \(h^{(1)}\) is defined on this pair by
\begin{equation}
    h^{(1)}v = u, \qquad h^{(1)}u=0.
\end{equation}
The full homotopy is then given by
\begin{equation} \label{eq: ht-prim}
     h = h^{(1)} + h^{(1)} \delta_Q^{(2)} h^{(1)} + (h^{(1)} \delta_Q^{(2)} h^{(1)})^2 h^{(1)} + \dots
\end{equation}
so that we have $\Psi = h S_{BV}$.

In the next subsection, we will realize the $Q_{max}$-exactness more directly, by using the realization of the maximal twist as a further twist of the minimal twist.

\begin{rmk}
    This result can be interpreted in terms of the pure spinor superfield formalism. Recall that the variety of square-zero elements is all of the affine space $S_+$. The IKKT multiplet is obtained in the pure spinor superfield formalism from functions on the pure spinor cone inside $S_+$. Viewing this ring as a sheaf over $S_+$, we see that it has trivial fibers at all the non-minimal twisting supercharges. On general grounds (see e.g. the discussion in~\cite{spinortwist,6dbundles}), one obtains twists that are locally trivial in field space in such situations. 
\end{rmk}

\subsubsection{The non-minimal twist as a further twist}
The non-minimal twist can be alternatively viewed as a further deformation of the minimal twist. Explicitly, after picking a minimal twisting supercharge, we can deform $\fp_{IKKT}$ to the dg Lie algebra $(\fp_{IKKT} , [Q_{\text{min}},-])$. It is easy to see that its cohomology is
\begin{equation}
    (\mathfrak{sl}(5) \ltimes \wedge^2 \CC^5) \oplus \Pi \CC^5 .
\end{equation}
Picking any non-zero element in the odd part $\Pi \CC^5$ defines a valid choice of deformation away from the minimal twist. Fixing such an element, induces a decomposition $\CC^5 \cong \CC^4 \oplus \CC$ and thereby breaks the $\mathrm{SU}(5)$-symmetry to $\mathrm{SU}(4)$. 

A short calculation shows that the residual symmetries after this further deformation are purely bosonic and take the form
\begin{equation}
    \mathfrak{sl}(4) \ltimes (\wedge^2 \CC^4 \oplus (\CC^4)^\vee \oplus \CC^4).
\end{equation}
Recall that the stabilizer of a maximal twisting supercharge is $\Spin(7) \ltimes \CC^8$. Recalling the branching rules of $\Spin(7)$ to $\mathrm{SU}(4)$, we immediately find a match. Explicitly, $\mathfrak{sl}(4) \oplus \wedge^2 \CC^4$ is identified with the adjoint representation and $(\CC^4)^\vee \oplus \CC^4$ with the eight-dimensional spinor representation of $\Spin(7)$.

We can decompose the field content of the minimally twisted theory under $\mathrm{SU}(4)$.
\begin{equation}
    Z^i \mapsto (X^a , Z) \qquad \psi^{(2)ij} \mapsto (\psi^{ab} , \psi^{a})
\end{equation}
where $a = 1 \dots 4$  is now and index for the fundamental representation of $\mathrm{SU}(4)$. 

The original supersymmetry transformations of the IKKT model induce an action on these fields by homotopy transfer. Another way to see this is by expanding the original IKKT action around the minimally twisted background. Then, we specialize these transformation to the further twisting supercharge. We find the following linear transfomation that render the twist acyclic as expected.
\begin{equation}
\begin{array}{ll}
    % Left Column & Right Column
    \delta_Q^{(1)} c = Z & \delta_Q^{(1)} Z^* = c^*  \\[6pt]
    \delta_Q^{(1)} Z = 0 & \delta_Q^{(1)} \bar{X}^{*a} = 0 \\[6pt]
    \delta_Q^{(1)} X^a = \psi^a & \delta_Q^{(1)} \psi^{*ab} = 0 \\[6pt]
    \delta_Q^{(1)} \psi^{ab} = \psi^{* ab} \qquad \qquad& \delta_Q^{(1)} \psi^{a} = X^{*a} \\[6pt]
    \delta_Q^{(1)} \psi^{a} = 0 & \delta_Q^{(1)} c^* = 0 \\
\end{array}
\end{equation}

From these considerations, one can obtain $S_{BV}$ in a $Q_{max}$-exact form as follows. The residual odd translations of the minimal twist act by the derivations
\begin{equation}
    \partial_i := \frac{\partial}{\partial \theta^i}, \qquad i=1,\ldots,5 .
\end{equation}
Choosing a nonzero residual odd translation gives a further twist.
After an \(\SU(5)\) rotation, we may take $q=\partial_5$. Thus the further-twisted model is governed by
the dg Lie algebra $\bigl(A \otimes \mathfrak{gl}_N,\; q=\partial_5, \;[-,-] \bigr)$. This complex is contractible, as computed above: multiplication by \(\theta^5\) gives an explicit homotopy $h=\theta^5$
satisfying
\begin{equation}
      qh+hq=1.
\end{equation}
In the BV formalism, the further twist adds the Hamiltonian for the vector field \(\delta_q\alpha=q\alpha\).  Thus the maximally twisted action in
this model is
\begin{equation}
  S_{\max}(\alpha) = \int d^5 \theta \mathcal{L}_{max} 
  =
  \frac{1}{2}
  \int d^5\theta\;
  \operatorname{Tr}\bigl(\alpha\,q\alpha\bigr)
  +
  \frac{1}{6}
  \int d^5\theta\;
  \operatorname{Tr}\bigl(\alpha[\alpha,\alpha]\bigr).
\end{equation}
Since \(\delta_q\alpha=q\alpha\), one has
\begin{equation}
  \delta_q \mathcal L_{\max}
  =
  \partial_5 \mathcal L_{\max}.
\end{equation}
Therefore we can write a primitive
\begin{equation}
  \Psi_{\max}
  :=
  \int d^5\theta\;
  \theta^5\,\mathcal L_{\max}(\alpha)
\end{equation}
which satisfies
\begin{equation}
  \delta_q\Psi_{\max}
  =
  \int d^5\theta\;
  \theta^5\partial_5\mathcal L_{\max}
  =
  \int d^5\theta\;\mathcal L_{\max}
  =
  S_{\max}.
\end{equation}
so $S_{\max}=\delta_q\Psi_{\max}$.
It may be useful to rewrite this in manifest \(SU(4)\) notation.
Decompose
\begin{equation}
  \alpha
  =
  \beta+\theta^5\gamma,
  \qquad
  \beta,\gamma\in
  \mathbb C[\theta^1,\ldots,\theta^4]\otimes\mathfrak g .
\end{equation}
Then
\begin{equation}
  \delta_q\beta=\gamma,
  \qquad
  \delta_q\gamma=0.
\end{equation}
The action and its primitive become
\begin{equation}
  S_{\max}
  =
  \int d^4\theta\;
  \operatorname{Tr}\left(
    \frac{1}{2}\gamma\gamma
    +
    \frac{1}{2}\gamma[\beta,\beta]
  \right),
\end{equation}
and
\begin{equation}
  \Psi_{\max}
  =
  \int d^4\theta\;
  \operatorname{Tr}\left(
    \frac{1}{2}\beta\gamma
    +
    \frac{1}{6}\beta[\beta,\beta]
  \right).
\end{equation}
We remark that this simple construction of the primitive works precisely because the further deformation from the minimal to the non-minimal twist acts linearly on the fields so that the higher order terms in~\eqref{eq: ht-prim} vanish.

\subsubsection{Comparison to twisted type IIB supergravity} \label{sec: ikkt-sugra-nonmin}
Twisting supercharges in the type IIB super Poincar\'e algebra have been recently classified in~\cite{iib-nv} and indeed, there is a $\Spin(7)$-invariant supercharge closely related to the non-minimal twisting supercharge in IKKT. Recall that the supertranslation algebra for type IIB supersymmetry is
\begin{equation}
    \Pi( S_+ \otimes \CC^2) \oplus V,
\end{equation}
where $V$ is the vector representation of $\Spin(10)$ and $\CC^2$ is the R-symmetry space. The bracket is given by the vector invariant $\Gamma^I$ together with a symmetric bilinear form $(-,-)_W$ on $\CC^2$. In order to produce a square-zero supercharge in the type IIB algebra, we simply take our non-minimal IKKT twisting supercharge $Q$ and form the tensor product $Q \otimes w$ with an isotropic vector $w \in \CC^2$.

To our knowledge, this $\Spin(7)$-invariant twist of type IIB supergravity has not been investigated in the literature yet. As explained in~\cite{iib-nv}, this twist features a single surviving translation in the residual supertranslation algebra so that one expects it to give a mixed topological-holomorphic theory on backgrounds of the type $\RR^8 \times \CC$. However, a careful investigation of the twists of branescan central extensions in this context shows that the fundamental string extension of the IIB supersymmetry algebra descends via homotopy transfer to a differential in the residual symmetries of the twisted theory and forms a acyclic pair with the last remaining translation. Since (twisted) supergravity is geometrically best viewed as a theory of deformations of Cartan connections (or other suitable geometric structures) modeled on the (residual) supersymmetry algebra, this acyclicity suggests that this twist of supergravity is perturbatively trivial, consistent with the above result on the perturbative triviality of the non-minimally twisted IKKT model. Further work on this perspective will appear in~\cite{twistedBranes}.

In addition, we expect the $\Spin(7)$-twist of type IIB supergravity to be closely related to a further deformation of BCOV theory by turning on a linear superpotential. Indeed viewing the non-minimal twist of IKKT as a further twist (see above), the choice of the twisting supercharge in the residual symmetries mirrors the choice of a linear superpotential for BCOV theory on $\CC^5$, the gravitational side of the duality.

\section{Twisted holography for BFSS matrix quantum mechanics}

\subsection{Review of the BFSS model}
The D0-brane matrix quantum mechanics (MQM) is commonly referred to as the BFSS theory, cf.~\cite{BFSS}. It can be obtained by dimensional reduction of ten-dimensional super Yang--Mills theory to one dimension. Therefore, its fields are a collection of nine bosonic matrices $X^i$ that transform in the vector representation of $\Spin(9)$ together with 16 fermions $\psi_\alpha$ transfoming in the spin representation and a gauge field $A$. All these fields are further valued in $\mathfrak{u}(N)$. The action is
\begin{equation}\label{eq: action}
    S_{BFSS} = \frac{1}{g^2} \int \d t \: \mathrm{tr}( \frac{1}{2} (D_t X^i)^2 + \frac{1}{2} \psi^\alpha D_t \psi_\alpha - \frac{1}{4} [X_i, X_j]^2 - \frac{1}{2} \gamma^i_{\alpha \beta} \psi^\alpha [\psi^\beta, X_i] ),
\end{equation}
where, $D_t = \partial_t + [A,-]$ denotes the covariant derivative with respect to the gauge field.

Via dimensional reduction from ten-dimensional super Yang--Mills theory, the fields inherit an action by the super Lie algebra $\fp_{BFSS} = (\mathfrak{so}(9) \oplus V_t) \oplus \Pi S_{9d}$, where $V_t$ is the one dimensional space of translations along the remaining direction and $S_{9d}$ denotes the 16-dimensional spinor representation of $\Spin(9)$. The bracket between two odd elements is simply the trace $\Sym^2(S_{9d}) \longrightarrow V_t$. As in our discussion for the IKKT model, these correspond to the dynamical supersymmetries that are our main focus in this work. These are supplemented by 16 additional kinematical supercharges as we will briefly comment below.

The dynamical supersymmetry transformations act on the fields as follows.
\begin{equation} \label{eq: susy-trans}
    \begin{split}
        \delta X^i &= \varepsilon^\alpha \gamma^i_{\alpha \beta} \psi^\beta \\
        \delta \psi_\alpha &= D_t X^i \gamma_i \varepsilon + \frac{1}{2} [X_i, X_j] \gamma^{ij}_{\alpha \beta} \varepsilon^\beta \\
        \delta A &= \varepsilon^\alpha \psi_\alpha
    \end{split}
\end{equation}
Working in the BV formalism, we introduce a ghost $c$ for gauge transformation as well as antifields and antighosts $(X^{*}_i, A^*, \psi_\alpha^*, c^*)$. The action~\eqref{eq: action} is supplemented by terms encoding gauge transformations,
\begin{equation}
    S_{\text{gauge}} = \int c^*[c,c] + A^* (\dot{c} + [c,A]) + X^{*i} [c,X_i] + \psi^* [c,\psi]
\end{equation}
as well as supersymmetry transformations
\begin{equation}
    S_{\text{susy}} =  \int X^*_i \delta X + \psi^* \delta \psi + A^* \delta A + c^* (\varepsilon \gamma_i \varepsilon X^i + \varepsilon^2 A) + (\varepsilon \gamma^i \varepsilon) \psi^* \gamma_i \psi^* - 2 (\varepsilon \psi^*)^2.
\end{equation}
As in the IKKT model, we observe a structure corresponding to closure terms of the supersymmetry algebra up to gauge transformations and the equations of motion. As before, we take the BV action to be the sum of these three terms with the BV differential splitting into terms corresponding to the equations of motion, gauge transformations, and supersymmetry transformations.

The full supersymmetry module structure of the BV fields can be read off from the BV action. We find:
\begin{equation}
\begin{split} \label{eq: susy-bfss}
& \delta c = (\varepsilon \gamma^i \varepsilon) X_i + \varepsilon^\alpha \varepsilon_\alpha A \\
& \delta X^i = \varepsilon \gamma^i \psi \\
& \delta \psi = D_t X^i \gamma_i \varepsilon + \frac{1}{2} [X_i, X_j] \gamma^{ij} \varepsilon + (\varepsilon \gamma^i \varepsilon)(\psi^* \gamma_i) - 2 \varepsilon (\varepsilon \cdot \psi^*) \\
& \delta A = \varepsilon \psi,
\end{split}    
\end{equation}
as well as for the antifields:
\begin{equation}
    \begin{split}
        &\delta X_i^* = \varepsilon \gamma_i D_t \psi + \gamma_{ij} [X^j, \psi] \varepsilon + (\varepsilon \gamma_i \varepsilon) c^* \\
        &\delta \psi^* = \varepsilon \gamma^i X_i^* + \varepsilon A^* \\
        &\delta A^* = \varepsilon^\alpha \varepsilon_\alpha c^* \\
        &\delta c^* = 0
    \end{split}
\end{equation}

\subsubsection{Relation to M-theory}
The BFSS MQM in the $N \rightarrow \infty$ limit conjecturally reconstructs the non-perturbative S-matrix of M-theory in eleven-dimensional flat spacetime~\cite{BFSS}. Arguments for this proposal were given in \cite{Seiberg:1997ad}, with tests and extensions in, e.g., \cite{Sen:1997we, Susskind:1997cw, Paban:1998ea, Kabat:1997sa, Taylor:1998tv}; see \cite{Taylor:2001vb} for a review. Evidence for this conjecture was further provided by the existence of solutions in the matrix model corresponding to M2 and M5 branes~\cite{Berkooz:1996is, Banks:1996nn, Kabat:1997im, Chepelev:1997vx}. More recently, a BPS-saturated M-theory three-point amplitude was computed from the MQM in \cite{HM}, and eleven-dimensional Lorentz invariance was recovered by analyzing soft theorems \cite{Tropper:2023fjr, Herderschee:2023bnc}.

Similar to our discussion above for the IKKT model, the 16 dynamical supercharges of the BFSS model are also supplemented by 16 additional kinematic supercharges combining to 32 supercharges consistent with the M-theoretic dual. Again these act as shift symmetries on the fermion, since they are Goldstinos for the broken supersymmetries in the presence of the D0 brane.
\subsection{Twisting supercharges}
The dynamical supersymmetries live in the 16-dimensional spinor representation $S_{9d}$ of $\Spin(9)$. The equation $[Q,Q] = 0$ is a single equation, so that the variety of square-zero elements is a 15-dimensional hypersurface in $S_{9d}$. Expanding $Q = \varepsilon^\alpha Q_{\alpha}$ into coordinates, the defining equation is simply
\begin{equation}
    \delta^{\alpha \beta} \varepsilon_\alpha \varepsilon_\beta = 0.
\end{equation}
The variety decomposes into two orbits:
\begin{itemize}
    \item[---] A minimal orbit of dimension 11. In this orbit, the stabilizer is $\mathrm{SU}(4) \ltimes (\wedge^2 \CC^4 \oplus \CC^4)$.
    \item[---] The open orbit of dimension 15. In this orbit, the stabilizer is $G_2 \ltimes \CC^7$. 
\end{itemize}
We note that the two different twisting supercharges are distinguished by their vector invariants. In the minimal orbit we have
\begin{equation}
    Q_{\text{min}}\gamma^i Q_{\text{min}} = 0 \qquad \text{for } i=1 \dots 9,
\end{equation}
while in the maximal orbit, we have $Q_{\text{max}}\gamma^i Q_{\text{max}} \neq 0$. The supercharge used in the localization computation of the recent work~\cite{Asano:2026udd} is a maximal twisting supercharge.

We further can compare this to the twisting supercharges for type IIA supergravity and M-theory. As we will discuss in~\S\ref{sec: compare-11d} in more detail, there are two distinct classes of twisting supercharges in the case of eleven-dimensional supersymmetry. The minimal twist is invariant under $\mathrm{SU}(5)$ and lives on backgrounds of the form $\CC^5 \times \RR$; the maximal twist is invariant under $\mathrm{SU}(2) \times G_2$ and yields a theory on backgrounds of type $\RR^7 \times \CC^2$. Both of these twists, can be dimensionally reduced along a holomorphic direction to give twists of type IIA. In the former case this is the $\mathrm{SU}(4)$-twist of type IIA; this dimensional reduction was described in~\cite{RSW}.

\subsection{The minimal twist of BFSS}
A description of the minimal twist can be again obtained in a straightforward way via dimensional reduction. Reducing holomorphic Chern-Simons theory to one dimension gives
\begin{equation}
    \Omega^\bullet(\RR) \otimes_{\RR} \CC[\theta^1, \dots , \theta^4] \otimes \mathfrak{gl}_N,
\end{equation}
where the $\theta^i$ are fermionic variables transforming in the fundamental representation of $\mathrm{SU}(4)$. Upon quasi-isomorphism we can identify this simply with $\CC[\theta^1, \dots , \theta^4] \otimes \mathfrak{gl}_N$. The BV action for $\alpha \in \Omega^\bullet(\RR) \otimes_{\RR} \CC[\theta^1, \dots , \theta^4] \otimes \mathfrak{gl}_N$ is
\begin{equation} \label{eq:action-bfss}
    S_{BV}[\alpha] = \int \d t \d \theta^5 \mathrm{tr} (\alpha \d_t \alpha + \frac{2}{3} \alpha [\alpha, \alpha]).
\end{equation}

\subsubsection{Direct computation of the minimal twist}
We start by decomposing the fields into representations of $\SU(4)$. This gives
\begin{equation}
    X^i \mapsto (Z^a, \bar{Z}_a, X) \qquad \psi_\alpha \mapsto (\psi^{(0)} , \psi^{(1)}, \dots , \psi^{(4)}),
\end{equation}
where $a=1 \dots 4$ and the superscript for the fermion denotes form degree according to the decomposition $S = \wedge^\bu (\CC^4)^\vee$.
Similarly for the antifields, we have
\begin{equation}
        X^{*i} \mapsto (Z_{a}^*, \bar{Z}^{*a}, X^*) \qquad \psi^*_\alpha \mapsto (\psi^{*(0)} , \psi^{*(1)}, \dots , \psi^{*(4)}),
\end{equation}
We proceed as in~\S\ref{sec: min-tw-ikkt} and specialize the parameters of the supersymmetry transformation to a minimal twisting supercharge by setting $\varepsilon^{(0)} =1$ and all other components to zero. The total BV differential now decomposes into terms organized both by polynomial degree in the BV fields as well as number of derivatives. Terms that are linear and without time derivatives arise only from $\delta_Q$ and lead to the formation of acyclic pairs and thereby cancelations in cohomology. Terms involving derivatives and non-linear contributions give rise to the BV differential of the twisted theory.

We decompose the linear supersymmetry transformations. We find for the BV fields:
\begin{equation}
\begin{array}{ll}
    % Left Column & Right Column
    \delta_Q^{(1)} c = 0 & \delta_Q^{(1)} \psi^{(0)} = \varepsilon_{abcd} \psi^{*(4)abcd} \\[6pt]
    \delta_Q^{(1)} Z^a = 0 & \delta_Q^{(1)} \psi^{(1)} = 0 \\[6pt]
    \delta_Q^{(1)} \bar{Z}_a = \varepsilon_{abcd} \psi^{(3)bcd} \qquad \qquad & \delta_Q^{(1)} \psi^{(2)} = 0 \\[6pt]
    \delta_Q^{(1)} X = \varepsilon_{abcd} \psi^{(4)abcd} & \delta_Q^{(1)} \psi^{(3)} = 0 \\[6pt]
    \delta_Q^{(1)} A = \varepsilon_{abcd} \psi^{(4)abcd} & \delta_Q^{(1)} \psi^{(4)} = 0 \\
\end{array}
\end{equation}
For the antifields, we have:
\begin{equation}
\begin{array}{ll}
    % Left Column & Right Column
    \delta_Q^{(1)} c^* = 0 & \delta_Q^{(1)} \psi^{*(0)} = X^* + A^* \\[6pt]
    \delta_Q^{(1)} Z^*_a = 0 & \delta_Q^{(1)} \psi^{*(1)a} = \bar{Z}^{*a} \\[6pt]
    \delta_Q^{(1)} \bar{Z}^{*a} = 0 \qquad \qquad & \delta_Q^{(1)} \psi^{*(2)} = 0 \\[6pt]
    \delta_Q^{(1)} X^* = 0 & \delta_Q^{(1)} \psi^{*(3)} = 0 \\[6pt]
    \delta_Q^{(1)} A^* = 0 & \delta_Q^{(1)} \psi^{*(4)} = 0 \\
\end{array}
\end{equation}
In total, we can read off the cohomology to find that
\begin{equation}
    c, X-A, Z^a, \psi^{(1)}, \psi^{(2)}, \psi^{*(2)}, \psi^{*(3)}, Z^{*a}, X^*-A^*, c^*
\end{equation}
are the BV fields of the twisted theory.

We identify these with the fields of dimensionally reduced holommorphically twisted super Yang--Mills theory along the following table
\begin{table}[h] 
\centering
\renewcommand{\arraystretch}{1.5} 
\begin{tabular}{|c|c|c|c|c|c|}
\hline
 & $(\theta^a)^0$ & $(\theta^a)^1$ & $(\theta^a)^2$ & $(\theta^a)^3$ & $(\theta^a)^4$ \\
\hline
$1$ & $c$ & $Z^a$ & $\psi^{(2)}$ & $\psi^{*(3)}$ & $(X^* - A^*)$ \\
\hline
$\d t$ & $(X-A)$ & $\psi^{(1)}$ & $\psi^{*(2)}$ & $Z^{* a}$ & $c^*$ \\
\hline
\end{tabular}
\caption{Fields of minimally twisted BFSS.} \label{table bfss}
\end{table}
Terms from gauge transformations, the equations of motions, and the derivative pieces of $\delta_Q$ induce the BV differential of the twisted theory. It is not difficult to see that these terms collectively give rise to the dg Lie structure on
\begin{equation}
    \left( \Omega^\bullet(\RR) \otimes \CC[\theta^1, \dots , \theta^4] \otimes \mathfrak{gl}_N \: , \: \mathrm{d}_{dR} \: , \: [-,-] \right)
\end{equation}
as we expect. For example, the gauge transformation acting on $X$ and $A$ induces a term
\begin{equation}
    \delta_{\mathrm{gauge}} (X-A) = D_t c
\end{equation}
corresponding to the de Rham differential acting on a zero-form in theta degree zero and the bracket between zero- and one-forms. Or, the derivative piece in the supersymmetry transformation acting on $\psi^{(1)}$
\begin{equation}
    \delta_Q \psi^{(1)a} = D_t Z^a,
\end{equation}
corresponding to $D_t$ acting on a zero-form in theta-degree one. Assembling all these pieces, it is not difficult to see that the BV action reproduces the dimensional reduction~\eqref{eq:action-bfss}.

\begin{rmk}
    As an alternative to the two-step procedure sketched above, one can also directly incorporate terms involving time derivatives and then perform a descent procedure. This makes the fact that the twisted theory assembles into differential forms even more apparent. Indeed, recall that in the descent procedure, descendants of an operator $\cO^{(0)}$ are constructed by the rule
    \begin{equation}
        \delta (\cO^{(1)}) = D_t \cO^{(0)} \d t.
    \end{equation}
    Terms like this originate from both gauge transformations and equations of motion as well as $\delta_Q$. This identifies the (linear local operators dual to the) second row of Table~\ref{table bfss} as the descendants of the first row. When placing the MQM on, say, a Euclidean circle, the twist necessarily contains in its observables supersymmetric Wilson loops (second row, first column), previously considered by \cite{MM}. 
\end{rmk}

\subsection{Twisted holography for minimally twisted BFSS} \label{sec: hol-bfss}
Running the same chain of arguments as in~\S\ref{sec: hol-ikkt}, we find that single trace operators in the minimally twisted BFSS model yield to polyvector fields describing BCOV theory on $\CC^4$. Including the two topological directions we find
\begin{equation}
    (\mathrm{PV}^{\bu, \bu}(\CC^4)[\![t]\!] \: , \: \bar{\partial} + t \partial_\Omega) \otimes (\Omega^\bu(\RR^2) , \d_{dR}).
\end{equation}
This theory can be connected to type IIA string theory and supergravity as follows. From the perspective of topological string theory, this is the target space theory originating from the B-model on $\CC^4$ times the A-model on $\RR^2$. It is conjectured to be the $\SU(4)$ twist of the type IIA string~\cite{CostelloLi}. We note that in this mixed A/B-model, the category of branes is a product between A-branes in $\RR^2$ (given by Lagrangian submanifolds in $\RR^2$) and B-branes on $\CC^4$ (given by coherent sheaves on $\CC^4$). Twisted D0 branes are of the form $\RR \times \{p\}$ where $p$ is a point in $\CC^4$. Indeed, the self-Ext of such objects precisely recovers the minimal twist of the BFSS model as described above.

In order to relate to supergravity, one again truncates to minimal BCOV theory described by the subcomplex
\begin{equation} \label{eq: min-bcov}
    \left( \bigoplus_{i+j \leq 3} t^i \mathrm{PV}^{j,\bullet}(\CC^4) \: , \: \bar{\partial} + t \partial_\Omega \right) \otimes (\Omega^\bu(\RR^2) , \d_{dR}).
\end{equation}
The $\mathrm{SU}(4)$-twist of IIA has been addressed in the supergravity approximation by viewing it as a dimensional reduction of the minimal twist of eleven-dimensional supergravity along a holomorphic direction~\cite{spinortwist, RSW}. The theory obtained by that procedure is related to~\eqref{eq: min-bcov} by choosing potentials for some of the field strengths. A description of the interactions can be found in~\cite[\S 5.2]{RSW}.

Many aspects discussed in~\S\ref{sec: hol-ikkt} for the IKKT model transfer to this setting. For example, the minimal model of minimal BCOV theory is identified with $\mathrm{SHO}(\CC^{4|4})$ leading to a similar discussion of infinite-dimensional symmetry algebras; furthermore, the difference between full and minimal BCOV theory is expected to encode the twists of the massive string modes of the IIA string.

\begin{rmk}
There is another notion of holography in this system arising by instead taking the 't Hooft limit of the BFSS MQM \cite{Itzhaki:1998dd}, which gives a more standard, (non-conformal) example of gauge-gravity duality. The closed string dual is well-described by the ten-dimensional black 0-brane geometry, or its eleven-dimensional uplift depending on the effective coupling (see e.g. \cite{Polchinski:1999br}). Since the twisted theory is a protected algebra that is insensitive to the continuous coupling $\lambda$, it is natural to expect that it witnesses a protected subsector common to all effective supergravity descriptions; indeed, we have seen that it is naturally sensitive to a twisted ten-dimensional dual which we will identify as a dimensional reduction of the putative twisted eleven-dimensional dual. To study twisted holography for the Maldacena duality, one must also incorporate D-brane backreaction on the closed string side of the duality, effected by adding a source term to the field equation and studying fluctuations around the resulting background. Unlike in the familiar examples of twisted AdS/CFT \cite{Costello:2018zrm, Costello:2020jbh} where the backreaction induces a Beltrami differential that deforms the complex structure, in this setting it does not correspond to a geometric deformation, so the twisted dual is harder to interpret. 
\end{rmk}

\subsection{Non-minimally twisted BFSS}
We now investigate the non-minimal twist of BFSS. Similar to the non-minimal twist of IKKT, we find that the deformed BV complex is acyclic, signaling cohomological triviality\footnote{As before, this is only true locally, i.e. in the neighborhood of a point in the moduli stack of classical solutions modulo gauge transformations.}. However, we find that the theory still has non-trivial content constructed from topological descent.

Choosing the twisting supercharge breaks the $\Spin(9)$ symmetry according to
\begin{equation}
    \Spin(9) \supset \Spin(7) \times \Spin(2) \supset G_2 \times \mathrm{U}(1).
\end{equation}
The spinor and vector representations decompose as
\begin{equation}
    \begin{split}
        S &\longrightarrow (1 \oplus V_7) \otimes (\textbf{1}^1 \oplus \textbf{1}^{-1}) \\
        V &\longrightarrow V_7 \oplus \textbf{1}^{+2} \oplus \textbf{1}^{-2},
    \end{split}
\end{equation}
where $V_7$ is the seven-dimensional representation of $G_2$.
We decompose the field content accordingly
\begin{equation}
    \begin{split}
        X &= (Z_{+2}, \bar{Z}_{-2} ,  X^A) \\
        \psi &= (\psi_+ , \psi_-, \psi_+^A , \psi_-^A),
    \end{split}
\end{equation}
where now $A=1 \dots 7$ is an index for $V_7$. We fix the twisting supercharge in the representation $1_{G_2} \otimes \textbf{1}^{-1}$. The linear supersymmetry transformations acting on fields decompose as follows.
\begin{equation}
    \begin{array}{ll}
        \delta_Q^{(1)} c = Z_{+2} & \delta_Q^{(1)} \psi_{+} = 0\\[6pt]
        \delta_Q^{(1)} Z_{+2} = 0 & \delta_Q^{(1)} \psi_{-} = \psi^*_{+} \\[6pt]
        \delta_Q^{(1)} \bar{Z}_{-2} = \psi_{-} & \delta_Q^{(1)} \psi^A_{+} = \psi^{*A}_{-} \\[6pt]
        \delta_Q^{(1)} X^A = \psi^A_{+} \qquad \qquad & \delta_Q^{(1)} \psi^A_{-} = 0 \\[6pt]
        \delta_Q^{(1)} A = \psi_{+}
    \end{array}
\end{equation}
For the antifields, we find:
\begin{equation}
\begin{array}{ll}
    % Left Column & Right Column
    \delta_Q^{(1)} Z^*_{-2} = c^* & \delta_Q^{(1)} \psi^*_{+} = Z^*_{+2}  \\[6pt]
    \delta_Q^{(1)} \bar{Z}^*_{+2} = 0 & \delta_Q^{(1)} \psi^*_{-} = A^*  \\[6pt]
    \delta_Q^{(1)} X^*_A = 0 & \delta_Q^{(1)} \psi^{*A}_{+} = 0 \\[6pt]
    \delta_Q^{(1)} A^* = 0 & \delta_Q^{(1)} \psi^{*A}_{-} = X^{*A} \\[6pt]
    \delta_Q^{(1)} c^* = 0
\end{array}
\end{equation}
Again, we see that this complex is completely acyclic. As in the IKKT model, this implies that the non-minimally twisted BFSS action is exact, $S_{BV} = Q \Psi$, provided we assume periodic boundary conditions (such as when placing the theory on a Euclidean circle $S^1$) so that the supersymmetry variations do not produce anomalous boundary terms. $\Psi$ can be constructed as described in~\S\ref{sec: spin7-ikkt}.

However, in analogy to Donaldson--Witten theory, we can still construct non-trivial observables by moving to the basic subcomplex (see for instance \S 10 of \cite{Cordes:1994fc} for an explanation of these notions) and performing descent. In practice, this can be done by removing the ghost $c$ from the complex and imposing gauge invariance by hand. Then we have closed operators of the form
\begin{equation}
    \mathrm{tr}(Z^n) \qquad \text{for } n=1 \dots N.
\end{equation}
Further, we note that the full transformation of $\psi_+$ is
\begin{equation}
    \delta_Q \psi_+ = D_t Z_{+2} 
\end{equation}
so that topological descent gives rise to observables
\begin{equation}
    \int_{S^1} \d t \mathrm{tr}(\psi_+ Z^n).
\end{equation} As is well-known, an integrated descendant observable in the basic subcomplex depends only on the homology class of its insertion locus in spacetime, because slightly deforming the integration manifold changes the operator by a $\delta_Q$-exact term, per the descent equations. Furthermore, correlation functions of these gauge invariant observables, via localization, compute pairings or intersection numbers of cohomology classes on the $\delta_Q$-fixed moduli space of fields modulo gauge.\footnote{It would be interesting to explore connections between such basic subcomplexes in maximal twists, and other twisted or topological holography proposals, such as \cite{BenettiGenolini:2017zmu}.}
\subsubsection{Relation to twisted type IIA supergravity}
The minimal twisting supercharge for the BFSS model can be lifted to the type IIA supersymmetry algebra in the following way. Recall that the supertranslation algebra for type IIA is
\begin{equation}
    \Pi(S_+ \oplus S_-) \oplus V
\end{equation}
with the bracket between odd elements $Q = q_+ + q_-$ defined by
\begin{equation}
    [Q,Q] = \Gamma_+(q_+ , q_+) + \Gamma_-(q_-, q_-),
\end{equation}
where $\Gamma_\pm : \Sym^2(S_\pm) \longrightarrow V$ is the vector invariant between two spinors. We can identify the non-minimal twisting supercharge of BFSS as a chiral spinor $q_+$ of $\Spin(10)$ with non-vanishing vector invariant. To extend this to a square-zero supercharge in the IIA algebra, we simply choose a spinor of opposite chirality $q_-$ with $\Gamma_+(q_+,q_+) = -\Gamma_-(q_-,q_-)$ and take $Q = q_+ + q_-$.

Analyzing the residual supertranslation algebra for this twist of type IIA, one again finds that all but one translation is made exact by the twisting supercharge so that the expected background of the theory is again $\RR^8 \times \C$. However, similar to our discussion in~\S\ref{sec: ikkt-sugra-nonmin}, it turns out that accounting for the fundamental string extension renders the residual supertranslation algebra acyclic and hence this twist of supergravity perturbatively trivial. We expect the holographic dual of this twist to arise as a deformation of BCOV theory on $\CC^4$ by a linear superpotential.

\section{Comparisons to eleven dimensions} \label{sec: compare-11d}
The BFSS conjecture states that the matrix quantum mechanics is equivalent to full eleven-dimensional M-theory. In that light, we expect that the match with twists of type IIA string theory discussed in the previous section enhances to a match with twisted M-theory.

Most of what is known about twisted M-theory stems from computations in the supergravity approximation. The supersymmetry algebra allows for two distinct twisting supercharges.
\begin{itemize}
    \item[---] The minimal twist lives on backgrounds of the type $\CC^5 \times \RR$. Supercharges of this type are stabilized by $\mathrm{SU}(5) \ltimes \wedge^2 \CC^5$.
    \item[---] The maximal twist lives on backgrounds of type $\CC^2 \times \RR^7$. Supercharges of this type are stabilized by $(G_2 \times \mathrm{SU}(2)) \ltimes (V_7 \otimes \C^2 \oplus \wedge^2 \CC^2)$.
\end{itemize}

\subsection{Comparing to minimally twisted eleven-dimensional supergravity}
The minimal twist of eleven-dimensional supergravity was investigated in the free limit in~\cite{spinortwist}, and interactions were described in~\cite{RSW}. The BV fields of this twist can be described by the following cochain complex.
\begin{equation} \label{eq: comp-min-tw}
    \left[ \begin{tikzcd}[column sep = small]
    (\Omega^{0,\bu}(\CC^5) ,\bar{\partial}) \arrow[r, "\partial"] & (\Omega^{1,\bu}(\CC^5) , \bar{\partial}) \\ 
    (\mathrm{PV}^{1,\bu}(\CC^5) , \bar{\partial}) \arrow[r, "\partial_{\Omega}"] & (\mathrm{PV}^{0,\bu}(\CC^5),\bar{\partial}) \\
    \end{tikzcd} \right] \otimes (\Omega^\bu(\R) ,\d)
\end{equation}
Interactions are discussed extensively in~\cite{RSW}.
The minimal model of this $L_\infty$ algebra is a central extension of the exceptional infinite dimensional super Lie algebra $E(5,10)$.

This theory can be dimensionally reduced either along the topological direction or a holomorphic direction. Dimensionally reducing along the topological direction gives rise to the $\mathrm{SU}(5)$-invariant twist of type IIA. More importantly for our purposes, the reduction along a holomorphic direction yields a description of the $\mathrm{SU}(4)$-invariant twist of IIA. To be precise, one finds a version of minimal BCOV theory on $\C^4$ times the de Rham complex on $\RR^2$ that is related to what we found in~\S\ref{sec: hol-bfss} by choosing potentials for the field strengths in BCOV. 

On the other hand, the BCOV theory on $\CC^5$ that we obtained from minimally twisted IKKT is not obtained from eleven-dimensional supergravity but matches a holomorphic twist of type IIB supergravity. The relation to the $\SU(4)$-twist of type IIA is provided by T-duality. This is best illustrated using the terminology of topological string theory. Recall that the holomorphic twist of type IIB string theory corresponds to the pure B-model topological string. On the other hand, the $\mathrm{SU}(4)$-twist of type IIA is given by a product of the B-model on $\CC^4$ and the A-model on the two remaining topological directions.

T-duality establishes an equivalence between the A-model on the cylinder $\RR \times S^1$ with the B-model on $\CC^\times$, hence relating the twists $\SU(4)$-twist of IIA with the $\SU(5)$-twist of type IIB. We remark that the full identification is non-perturbative and cannot wholly be seen in the supergravity approximation (see~\cite{SuryaYoo} for a discussion in the context similar to this work).

This is consistent with IKKT and BFSS describing the dynamics of D(-1) and D0 branes respectively: Indeed, performing T-duality on a circle transverse to a D(-1)-brane transforms it into a Euclidean D0-brane wrapping the T-dual circle. The following diagram summarizes the relations between the minimal twist of the matrix models and string theories.
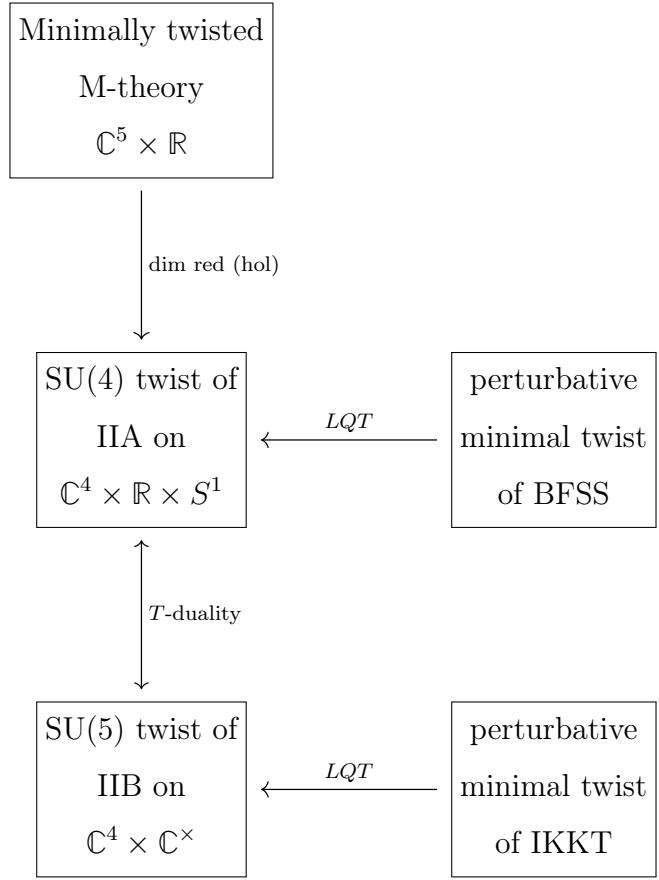
\begin{center}
    % We increase row/column separation to accommodate the larger boxes
    \begin{tikzcd}[row sep=2cm, column sep=2cm]
        % --- Row 1 ---
        \boxed{\begin{gathered} 
            \text{Minimally twisted} \\ 
            \text{M-theory} \\ 
            \mathbb{C}^5 \times \mathbb{R} 
        \end{gathered}}
        \arrow[d, "\text{dim red (hol)}"]
        & 
        \\
        % --- Row 2 ---
        \boxed{\begin{gathered} 
            \mathrm{SU}(4) \text{ twist of} \\ 
            \text{IIA on} \\ 
            \mathbb{C}^4 \times \mathbb{R} \times S^1 
        \end{gathered}}
        \arrow[d, <->, "T\text{-duality}"]
        & 
        \boxed{\begin{gathered}
            \text{perturbative} \\
            \text{minimal twist} \\ 
            \text{of BFSS} 
        \end{gathered}}
        \arrow[l, "LQT"']
        \\
        % --- Row 3 ---
        \boxed{\begin{gathered} 
            \mathrm{SU(5)} \text{ twist of} \\ 
            \text{IIB on} \\ 
            \mathbb{C}^4 \times \mathbb{C}^\times \end{gathered}}
        & 
        \boxed{\begin{gathered} 
            \text{perturbative} \\
            \text{minimal twist} \\ 
            \text{of IKKT} 
        \end{gathered}}
        \arrow[l, "LQT"']
    \end{tikzcd}
\end{center}

\subsection{Comparing to maximally twisted supergravity}
This twist was conjectured to be described by Poisson--Chern Simons theory by Costello~\cite{CostelloMtheory2}. This was verified in the free limit in~\cite{MaxTwist} and including interactions in~\cite{CY2}.

Explicitly the BV fields of Poisson--Chern--Simons theory on $\RR^7 \times \CC^2$ form the dg Lie algebra
\begin{equation}
    \left( \Omega^{(0,\bullet)}(\CC^2) \otimes \Omega^\bullet(\RR^7) \: , \: \bar{\partial}_{\CC^2} + \d_{\RR^7} \: , \: \{-,-\} \right),
\end{equation}
where $\{-,-\}$ is the Poisson bracket on $\CC^2$ extended to forms on $\R^7 \times \C^2$ by the wedge product. The BV action is
\begin{equation}
	S_{BV}(\alpha) = \int \Omega \wedge \alpha \left( \frac{1}{2}(\bar{\partial} + \d ) \alpha + \frac{1}{6} \{ \alpha , \alpha \} \right) ,
\end{equation}
where $\Omega = \d z_1 \wedge \d z_2$ is the holomorphic volume form.

In order to relate to twisted type IIA supergravity, we place the theory on $\RR^7 \times \CC \times \RR \times S^1$, where we think of the cylinder as being equipped with the standard complex structure. Then we perform dimensional reduction along the circle. This results in a theory on $\RR^8 \times \CC$ that can be identified with a product of de Rham forms in the eight topological directions together with BCOV theory with a linear superpotential in the complex direction (see~\cite[Remark 4.6]{SuryaYoo}). The linear superpotential deformation renders the BV complex acyclic, consistent with our observations on the non-minimal twist of BFSS.

% Bibliography

%% [A] Recommended: using JHEP.bst file
\bibliographystyle{JHEP}
\bibliography{generic.bib}

%% or
%% [B] Manual formatting (see below)
%% (i) We suggest to always provide author, title and journal data or doi:
%% in short all the informations that clearly identify a document.
%% (ii) please avoid comments such as "For a review'', "For some examples",
%% "and references therein" or move them in the text. In general, please leave only references in the bibliography and move all
%% accessory text in footnotes.
%% (iii) Also, please have only one work for each \bibitem.

\end{document}